\documentclass[aps,prl,preprint,groupedaddress]{revtex4-2}
\usepackage[dvipdfmx]{graphicx}
\usepackage{bm}
\usepackage{latexsym}
\usepackage{siunitx}
\usepackage{float}
\usepackage{tikz}
\usepackage{circuitikz}
\begin{document}

%\preprint{}

\title{Berry connection from many-body wave functions and superconductivity: Circuit quantization for superconducting qubits and absence of a dissipative quantum phase transition in Josephson junctions}


\author{Hiroyasu Koizumi}
%\thanks{}
\affiliation{Division of Quantum Condensed Matter Physics, Center for Computational Sciences, University of Tsukuba,Tsukuba, Ibaraki 305-8577, Japan}


\date{\today}

\begin{abstract}
The new superconductivity theory that attributes the $U(1)$ superconductivity phase to a Berry phase arising from many-body wave functions is applied to the circuit quantization for superconducting qubits.
 The phase-charge duality required for the occurrence of superconductivity in the standard theory becomes irrelevant in the new theory, and the absence of a dissipative quantum phase transition in Josephson junctions is explained. It is shown that a charge-decaying term leads to the compact phase description, and the appearance of Shapiro steps is explained without introducing normal current. 
\end{abstract}

%\keywords{}

\maketitle


%\subsection{Introduction}

The current standard theory of superconductivity is the one based on the BCS theory \cite{BCS1957}.
It has been a successful theory providing the way to calculate the superconducting transition temperature, explaining the energy gap formation, and predicting Josephson effects; however, it has not been successful in elucidating the high temperature superconductivity found in cuprates \cite{Muller1986}. 
A notable point of the cuprate superconductivity is that the superconducting transition temperature for the optimally-doped sample is not the pairing energy gap formation temperature, but the stabilization temperature for loop currents of the superconducting coherence length size \cite{Kivelson95}. This suggests that
the fundamental ingredient for superconductivity may actually be something that enables the appearance of the coherence-length-sized loop currents instead of the electron-pairing.

It is also notable that reexaminations of superconductivity from fundamental levels have
revealed two solid experimental facts pointing to the need for fundamental revisions in the standard theory \cite{Hirsch2009,Koizumi2011}.

One of them is the reversible phase transitions between normal and superconducting phases in the $H$-$T$ plane (for Type I superconductors)\cite{Hirsch2017,Hirsch2018,Hirsch2020,koizumi2020b}. 
A series of work \cite{Keesom1934a,Keesom1934b,Keesom1937,Keesom} indicate that the superconducting-normal state transition in the presence of a magnetic field occurs without energy dissipation; however, the supercurrent generated by the flow of electron pairs in the magnetic field inevitably produces the Joule heat during the superconducting to normal phase transition due to the existence of a significant number of broken pairs that flow with dissipation. 
 
The other is the mass of the electron in the London moment \cite{Hirsch2013b,koizumi2021}. 
The London moment has been measured in many different materials, ranging from the conventional superconductor \cite{Hildebrandt1964,Zimmerman1965,Brickman1969,Tate1989,Tate1990} to the high T$_{\rm c}$ cuprates \cite{VERHEIJEN1990a,Verheijen1990} and heavy fermion superconductors \cite{Sanzari1996}; the result always indicates that the mass $m$ in the London magnetic field should be the free electron mass $m_e$ if the electron charge $q=-e$ is employed; however,
the standard theory predicts it to be an effective mass $m^{\ast}$, as is evidenced by the
Gor'kov's derivation of the Ginzburg-Landau equations form the BCS one \cite{Gorkov1959}.  

The resolution for the above two discrepancies is provided by a new theory of superfluid that attributes the superfluidity to the appearance of  the nontrivial Berry connection from many-body wave functions \cite{koizumi2019,koizumi2021}. In this theory, the supercurrent is a topologically protected current generated by the collective mode created by the nontrivial Berry connection. 
In other word, a vector potential from the Berry connection exists in superconductors in addition to the
vector potential from electromagnetic field, and this non-trivial Berry connection is the source of the coherence-length-size loop currents. The success of the BCS theory is explained as due to the fact that this non-trivial Berry connection is stabilized by the electron-pair formation.

A salient feature of the new theory is that it is formulated in a particle number conserving way \cite{koizumi2019}; thus, the phase-charge duality is not necessary for the occurrence of superconductivity.
 The relevance of the phase-charge duality has bee believed to be evidenced in
the dissipative quantum phase transition in Josephson junctions \cite{Schmid1983,PhysRevX2021b}. However, its absence has been recently revealed  in the state-of-art experiment \cite{PhysRevX2021a,PhysRevX2021c}.

In the present work, we will explain the absence of the dissipative quantum phase transition in Josephson junctions based on the new superconductivity theory. 
We also present a new method for the circuit quantization for superconducting qubits,
and introduces a charge-decaying term for the Josephson junction. The circuit quantization is shown to be  almost identical to the standard one. 
The charge-decaying term leads to the compact phase description, and explains the appearance of Shapiro steps without introducing normal current.
 
% \subsection{Lagrangians for circuit elements}

We first, consider the quantization of circuits containing Josephson junctions \cite{Devoret1997,Nori2017}.
Usually, 
node flux
\begin{eqnarray}
\Phi_n = \int^t_{-\infty} V_n(t') dt'
\label{eq1}
\end{eqnarray}
where $V_n$ is node voltage is used as the dynamical variable.

Thus, the voltage between nodes $n$ and $m$ is given by
$\dot{\Phi_n} -\dot{\Phi_m}$.

The Lagrangian for a capacitor with capacitance $C$ existing between node $n$ and node $m$ is given by
\begin{eqnarray}
{\cal L}_C=
{ 1 \over 2}C \left(\dot{\Phi}_n-\dot{\Phi}_m \right)^2
\end{eqnarray}

In the new formalism, the following is used where a superconductor exists between nodes $n$ and $m$ 
\begin{eqnarray}
\Phi_n -\Phi_m= \int^{{\bf r}_n}_{{\bf r}_n}
{1 \over 2} \left[  {\bf A}^{\rm em}({\bf r}, t) -{{\hbar  } \over {2e}} \nabla \chi({\bf r}, t) \right] \cdot d {\bf r}
\label{eq3}
\end{eqnarray}
where ${\bf A}^{\rm em}$ is the vector potential for electromagnetic field, and $\chi$ is a $U(1)$ phase variable arising as the Berry phase from many-body wave functions.

The reason for the use of Eq.~(\ref{eq3}) is as follows: the time-derivative of the flux difference is calculated as
\begin{eqnarray}
\dot{\Phi}_n - \dot{\Phi}_m=-{1 \over 2}\int^{{\bf r}_n}_{{\bf r}_m}
\left[  {\bf E}^{\rm em}({\bf r}, t)+ \nabla \left( \varphi^{\rm em} +{{\hbar  } \over {2e}} \partial_t \chi({\bf r}, t) 
\right) \right] \cdot d {\bf r}
\label{eq4}
\end{eqnarray}
where
$
{\bf E}^{\rm em}=-\partial_t {\bf A}^{\rm em}-\nabla \varphi^{\rm em}
$
is used. Here ${\bf E}^{\rm em}$ is the electric field, and $\varphi^{\rm em}$ is the scalar potential for electromagnetic field.

The first term in Eq.~(\ref{eq4})
\begin{eqnarray}
-\int^{{\bf r}_n}_{{\bf r}_m}{\bf E}^{\rm em}({\bf r}, t)\cdot d {\bf r}
\label{eq5}
\end{eqnarray}
is the voltage due to the electric field. 

In order to interpret the second term in Eq.~(\ref{eq4}), 
%\begin{eqnarray}
%-\int^{{\bf r}_n}_{{\bf r}_m} \nabla \left( \varphi^{\rm em} +{{\hbar  } \over {2e}} \partial_t \chi({\bf r}, t) 
%\right) \cdot d {\bf r}
%\end{eqnarray}
we need to identify $\varphi^{\rm em} +{{\hbar  } \over {2e}} \partial_t \chi$.

Actually, it is the time-component partner of the gauge invariant vector potential
\begin{eqnarray}
{\bf A}^{\rm eff}={\bf A}^{\rm em} -{{\hbar  } \over {2e}} \nabla \chi
\end{eqnarray}
given by
\begin{eqnarray}
\varphi^{\rm eff}=\varphi^{\rm em} +{{\hbar  } \over {2e}} \partial_t \chi
\end{eqnarray}

It is related to the chemical potential $\mu$ as
\begin{eqnarray}
\mu=e\varphi^{\rm eff}=e\left(\varphi^{\rm em} +{{\hbar  } \over {2e}} \partial_t \chi \right)
\end{eqnarray}

Thus, the second term  in Eq.~(\ref{eq4}) is written as
\begin{eqnarray}
-\int^{{\bf r}_n}_{{\bf r}_m} \nabla \left( \varphi^{\rm em} +{{\hbar  } \over {2e}} \partial_t \chi({\bf r}, t) 
\right) \cdot d {\bf r}=
-{ 1 \over e}(\mu_n -\mu_m)
\label{eq9}
\end{eqnarray}

For the Josephson junction where the charging effect exists, the
balance between the voltage from the electric field and chemical potential difference is established as for a capacitor.
Thus, we have the following relations
\begin{eqnarray}
V=-\int^{{\bf r}_n}_{{\bf r}_m} {\bf E}^{\rm em}({\bf r}, t)\cdot d {\bf r}=-{1 \over e} \int^{{\bf r}_n}_{{\bf r}_m} \nabla \mu \cdot d {\bf r},
\label{eq18}
\end{eqnarray}
where $V$ is the voltage across the junction. Due to the two contributions each giving $V$, Eq.~(\ref{eq4}) is equal to $V$.

 We employ the following for the current through a Josephson junction that exists between nodes $n$ and $m$
 \begin{eqnarray}
J=J_c \sin \phi
\end{eqnarray}
where $\phi$ is the gauge invariant phase difference between the two superconductors in the junction.

In order to have the Josephson relation,
\begin{eqnarray}
{ {d \phi} \over {dt}}={{2eV} \over \hbar}
\label{eq19}
\end{eqnarray}
we adopt the following for $\phi$
\begin{eqnarray}
\phi={ e \over {\hbar } }\int_{{\bf r}_n}^{{\bf r}_m} \left[{\bf A}^{\rm em} -{{\hbar  } \over {2e}} \nabla \chi \right]
={ {2e} \over {\hbar } }\left[ {\Phi}_m-{\Phi}_n\right]
\label{eq-phi}
\end{eqnarray}
In the standard theory, the factor in front of the integral is ${ {2e} \over \hbar }$, instead of ${ {e} \over \hbar }$. Actually, in the Josephson's derivation, one of the contributions in Eq.~(\ref{eq4}) is missing \cite{Josephson62}.
We will come back to this point, later.

The Lagrangian for the Josephson junction ${\cal L}_J$ can be obtained from the condition
\begin{eqnarray}
J_c \sin \phi=  {{\partial {\cal L}_J} \over {\partial (\int_{{\bf r}_m}^{{\bf r}_n}{\bf A}^{\rm em} \cdot d{\bf r})}}
\end{eqnarray}

Neglecting the constant term, and adding the term that describes the charging, we use the following ${\cal L}_J$
\begin{eqnarray}
{\cal L}_J=E_J \cos { {2e} \over {\hbar } }\left( {\Phi}_m-{\Phi}_n\right)+{C_J \over 2}(\dot{\Phi}_n - \dot{\Phi}_m)^2,
\end{eqnarray}
where $E_J={ {\hbar  J_c} \over {e} }$.
Note that this $E_J$ is different from the standard one given by ${ {\hbar  J_c} \over {2e} }$.

%\subsection{Circuit quantization for a Cooper-pair box}

\begin{figure}[H]
 % \begin{center}
%   \begin{circuitikz}[american]
% \draw (0,0)   node[ground]{} node[left]{$1$}  
 %       to  [barrier=$E_J$,-*] (0,2) node[above]{$3$} 
 %    to[short] (1,2)
 %   to [C=$C_J$] (1,0)
%      to[short,-*] (0,0);
 %     \draw (1,2) 
 %     to[short] (2,2)
 %       to [C=$C_g$,-*] (4,2) node[above]{$2$} 
 %     to[V=$V_g$] (4,0)
 %     to[short] (1,0);
%    \end{circuitikz}
%  \end{center}
\includegraphics[width=5.0cm]{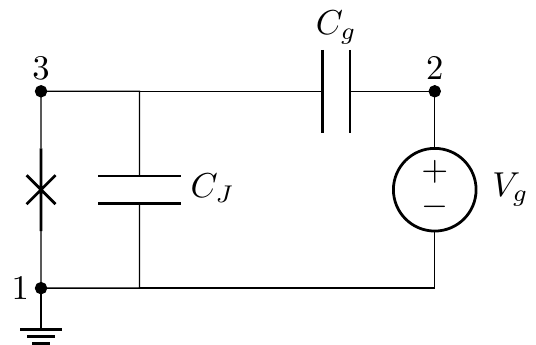}
  \caption{Cooper pair box. The Josephson junction is composed of components denoted as $E_J$ and $C_J$.
  $C_g$ denotes the gate capacitor, and $V_g$ supplies the gate voltage.}
  \label{CPB}
\end{figure}

Let us consider the Cooper-pair box in Fig.~\ref{CPB}. Its Lagrangian is given by
\begin{eqnarray}
{\cal L}_{CPB}={1 \over 2}C_g(\dot{\Phi}_3-\dot{\Phi}_2)^2+{1 \over 2}C_J(\dot{\Phi}_3-\dot{\Phi}_1)^2
+E_J \cos { {2e} \over {\hbar } }\left( {\Phi}_3-{\Phi}_1\right) 
\end{eqnarray}

The conjugate momentum for $\Phi_3$ is charge at node $3$, $Q_3=C_g(\dot{\Phi}_3-\dot{\Phi}_2)+C_J(\dot{\Phi}_3-\dot{\Phi}_1)$.

Using $\dot{\Phi}_2-\dot{\Phi}_1=V_g$, and neglecting a constant term, the Hamiltonian is obtained as
\begin{eqnarray}
{\cal H}_{CPB}={1 \over {2(C_g+C_J)}} (Q-C_g V_g)^2
-E_J \cos { {2e} \over {\hbar } } {\Phi}
\end{eqnarray}
where $\Phi_3-\Phi_1$ and $Q_3$ are replaced by $\Phi$ and $Q$, respectively.

We introduce $\hat{n}$ and $\hat{\phi}$ by
\begin{eqnarray}
\hat{n}=-{Q \over {2e}}, \quad \hat{\phi}= { {2e} \over \hbar }\Phi
\end{eqnarray}

The canonical quantization condition gives the following relation,
\begin{eqnarray}
[\hat{n}, \hat{\phi}]=i
\label{eq32}
\end{eqnarray}

The Hamiltonian is written as
\begin{eqnarray}
{\cal H}_{CPB}=E_C (\hat{n}-n_g)^2
-E_J \cos \hat{\phi}
\label{eq36}
\end{eqnarray}
where
$E_C ={(2e)^2 \over {2(C_g+C_J)}}$ and $n_g={{C_g V_g} \over {2e}}$.

Qubit states are obtained as the eigenstates of this Hamiltonian.
This is the same expression for the Hamiltonian based on the standard theory
except the $E_J$ value.

% \subsection{The duality principle for phase and charge in superconductivity}
 
 Although the Hamiltonian in Eq.~(\ref{eq36}) is almost the same as the one in the standard theory,
 there exists a significant difference;
the occurrence of  superconductivity is already taken into account before the quantization. 
The number operator $\hat{n}$ is not directly related by the number of Cooper-pairs.
This point is clearly seen in Eq.~(\ref{eq-phi}), showing that the factor in front of the integral is ${e \over \hbar}$, not ${{2e} \over \hbar}$.

 The occurrence of superconductivity is due to the presence of nontrivial $\nabla \chi$ in Eq.~(\ref{eq3}),
 and the duality arising from Eq.~(\ref{eq32}) is nothing to do with the occurrence of superconductivity.

In order to see why this difference between the present theory and the standard theory happened, we revisit the development of the Josephson effect theory.

 Josephson derived the Josephson relation based on the BCS theory and the tunneling current formalism by Cohen, Falicov, and Phillips (the CFP formalism)\cite{Cohen1962}. 
 The BCS superconducting state is given as a Cooper-pair number fluctuating state, and
 the CFP formalism includes only the voltage term from
 the chemical potential difference. 
 As a consequence he derived Eq.~(\ref{eq19}) by including one-half contributions in Eq.~(\ref{eq4}) (one in Eq.~(\ref{eq9})), without including the voltage from the electric field (one in Eq.~(\ref{eq5})), with employing the Cooper-pair charge $2e$.
 
 Next, Ferrell and Prange obtained the Josephson relation in Eq.~(\ref{eq19}) using the superconducting state constructed as a coherent sum of different particle number states \cite{Ferrell1963}. In their derivation, the Josephson relation arises from the fact that the phase and charge are canonical conjugate variables. Thus, duality principle for charge and phase came into the superconductivity theory.
 The dissipative quantum phase transition in Josephson Junction predicted was predicted based on this principle \cite{Schmid1983,PhysRevX2021b}. 
 
 Actually, the Josephson's derivation of the Josephson effect is defective.
 In contrast to the situation considered in Ref.~\cite{Cohen1962}, the charging effect is important in the Josephson junction, thus, the term for the voltage from the electric field must be included. 
It is also problematic that his derivation was not strictly gauge invariant as seen Eq.~(37) in Ref.~\cite{Josephson1969}, which should be compared with the gauge invariant expression in Eq.~(\ref{eq4}) in the present work.

Although the derived Josephson relation in Eq.~(\ref{eq19}) is correct, the derivation for it in the standard theory is defective. 

%\subsection{Absence of a dissipative quantum phase transition in Josephson junctions}

Now, we consider the absence of the dissipative quantum phase transition in Josephson junctions.

Let us consider the Hamiltonian
\begin{eqnarray}
{\cal H}={1 \over {2}} q^2
-J_c \cos \varphi -\varphi I
\end{eqnarray}
where $I$ is the current bias, and $q$ and $\varphi$ are two canonical conjugate variables \cite{PhysRevX2021b}.
This equation first seems to appear in Ref.~\cite{Anderson64}.
This can be considered as the Hamiltonian for a  particle in a potential energy $-J_c \cos \varphi -\varphi I$, which has a ``washboard potential tilted by $I$".

The Hamiltonian equation yields the equation for the charge conservation
\begin{eqnarray}
\dot{q}
+J_c \sin \varphi -I=0
\end{eqnarray}

The quantum mechanical state for ${\cal H}$ will be extended over $\varphi$ including values in different branches (the extended-$\varphi$) due to the tunneling.
In the standard theory, the delocalized $\varphi$ means the destruction of superconductivity according to the duality principle for phase and charge. Thus,
the state in the limit $I \rightarrow 0$ will be an insulator.

However, ${\cal H}$ with $I=0$ is essentially equivalent to ${\cal H}_{CPB}$. Then, 
the system should be superconducting that gives rise to qubit states. The experiment indicates that 
 the compact-$\varphi$ description is appropriate, and the dissipative quantum phase transition does not happen \cite{PhysRevX2021a}.

The above discrepancy is resolved in the new theory in which the origin of the $U(1)$ phase variable in superconductors is attributed to the Berry phase. Then, the existence of non-trivial 
\begin{eqnarray}
\langle \Phi \rangle=\int^{{\bf r}_3}_{{\bf r}_1}
{1 \over 2} \left\langle  {\bf A}^{\rm em}({\bf r}, t) -{{\hbar  } \over {2e}} \nabla \chi({\bf r}, t)  \right\rangle \cdot d {\bf r}
\end{eqnarray}
produces the supercurrent,
where $\langle \hat{O} \rangle$ denotes the expectation value of the operator $\hat{O}$.

We consider the absence of the dissipative phase transition using the modified version of ${\cal H}_{CPB}$ given by
\begin{eqnarray}
{\cal H}^m_{CPB}={1 \over {2C_J}} Q^2
-E_J \cos { {2e} \over {\hbar } } {\Phi}-I\Phi
\end{eqnarray}
The last term is added to include the current bias $I$.

From the Hamiltonian equation, the equation for the charge conservation
$\dot{Q}
+J_c \sin { {2e} \over {\hbar } } {\Phi} =I
$
is obtained. 
In the quantized case, it is approximated as
\begin{eqnarray}
\dot{\langle Q \rangle}
+J_c \sin { {2e} \over {\hbar } } \langle {\Phi}\rangle =I
\end{eqnarray}
In the new theory, the current here is superconducting one at each instant, even if $\dot{\langle Q \rangle}$ and  $\langle {\Phi}\rangle$ are time-dependent, and  $\langle {\Phi}\rangle$ extends over different branches. 
Thus, the system approached by $I \rightarrow 0$ is also superconducting in accordance with
the experimental result \cite{PhysRevX2021a}.

%\subsection{Shapiro steps}
The experiment also indicates that the compact-$\Phi$ description is appropriate \cite{PhysRevX2021a}.
We will show that the compact-${\Phi}$ is realized if we include the charge decaying effect due to the pair annihilation of opposite charges during the tunneling.
Actually, it also explains the Shapiro steps \cite{Shapiro63}.

By including the charge decaying effect, the Hamiltonian becomes
\begin{eqnarray}
{\cal H}_{SS}={1 \over {2C_J}} Q^2
-E_J \cos { {2e} \over {\hbar } } {\Phi}+\alpha Q \Phi+{ {C_J\alpha^2} \over 2}\Phi^2-I\Phi
\end{eqnarray}
where $\alpha Q \Phi$ describes the charge decaying effect due to the pair annihilation of opposite charges during the tunneling.

The term ${ {C_J\alpha^2} \over 2}\Phi^2$ arises to have the bias current equal to $I$.
This term makes $\Phi$ around $\Phi=0$ energetically favorable, thus leads to the
compact-$\Phi$ description. 
The presence of the charge decaying effect may be the reason for the appropriateness of the use of 
the compact-$\Phi$ indicated by the experiment \cite{PhysRevX2021a}.

The appearance of the Shapiro steps may be taken as the evidence for the presence of the charge decaying term. 
Let us derive a consequence of the charge decaying effect.

One of the Hamilton equations of the motion for ${\cal H}_{SS}$ yields 
\begin{eqnarray}
\dot{Q} 
+J_c \sin { {2e} \over {\hbar } }  {\Phi} +\alpha  Q +{ {C_J\alpha^2} }\Phi=I
\label{eq48}
\end{eqnarray}

The other Hamilton equations of the motion yields,
\begin{eqnarray}
\dot{\Phi} ={ {Q } \over C_J}+\alpha \Phi=V
\label{eq50}
\end{eqnarray}

The Shapiro steps appear when the voltage is given by
$
\dot{\Phi}=V_0 + V_1 \cos \omega_1 t 
$.
Thus, the charge is given by 
$
Q=C_J(V_0 + V_1 \cos \omega_1 t )-C_J \alpha \Phi
$, and its expectation value is
$
\langle Q \rangle =V_0 C_J-C_J \alpha \langle \Phi \rangle$. We also have $ \langle \dot{\Phi} \rangle=V_0$ and
$\Phi=V_0t+{V_1 \over \omega_1}\sin \omega_1+{\hbar \over  {2e} }\gamma_0$, where $\gamma_0$ is a constant. 

Using the generating function of Bessel functions,
$
e^{{x \over 2}(t-t^{-1})} =\sum_n J_n(x)t^n
$
 the following well-known result \cite{TinkhamText} is obtained
\begin{eqnarray}
J_c \sin { {2e} \over {\hbar } } \left( V_0t+{V_1 \over \omega_1}\sin \omega_1+{\hbar \over  {2e} }\gamma_0 \right) 
=J_c \sum_n (-1)^n J_n \left({{2eV_1} \over {\hbar \omega_1}} \right) \sin( \gamma_0 +\omega_0t -n\omega_1 t)
\label{eq53}
\end{eqnarray}
where $\omega_0={{2eV_0} \over \hbar}$.

 The stationary condition $\omega_0=n\omega_1$ yields
the step values of for the voltage $V_0$, 
\begin{eqnarray}
V_n={{\hbar \omega_1} \over {2e}}n
\label{eq54}
\end{eqnarray}

From Eqs.~(\ref{eq48}), (\ref{eq53}), and (\ref{eq54}), 
the current $I=I_n$ for the voltage step $V_n$ is obtained as
\begin{eqnarray}
{V_n \alpha C_J} -
J_c J_n \left({{2eV_1} \over {\hbar \omega_1}} \right) \leq I_n \leq {V_n \alpha C_J} +
J_c J_n \left({{2eV_1} \over {\hbar \omega_1}} \right)
\label{eq59}
\end{eqnarray}
where the variation of $\gamma_0$ values is taken into account. The equation (\ref{eq59}) explains the quantized voltage plateaus known as the {\em Shapiro steps} \cite{Shapiro63}.
Usually, ${V_n \alpha C_J}$ is considered as the normal current contribution using a resistance $R$ as $V_n /R$.
However it arises from supercurrent due to the presence of charging and charge-decaying here.

%In the standard theory, the above corresponds to
%\begin{eqnarray}
% \Delta({\bf r})\approx g e^{-i{\chi}({\bf r})} \langle 
%\hat{\Psi}_{\uparrow}({\bf r}) \hat{\Psi}_{\downarrow} ({\bf r'}) \rangle_{\rm BCS}
%\end{eqnarray}
%where $ \langle  \cdots \rangle_{\rm BCS}$ denotes that the expectation value is calculated using the particle number non-conserving state vector of the BCS theory.
%The supercurrent is due to this $e^{-i{\chi}({\bf r})}$. It appears due to the use of the particle number non-conserving formalism.
   
%\subsection{Conclusion}
In conclusion, 
we have presented the method for the circuit quantization for superconducting qubits using the new superconductivity theory, and explained the absence of the dissipative quantum phase transition in Josephson junctions as due to the irrelevance of the phase-charge duality for the occurrence of superconductivity. 
The charge decaying effect leads to the compact-$\Phi$ description, and explains the Shapiro steps without normal current.
We expect that the new circuit quantization method for superconducting qubits will contribute to the improvement of superconducting qubits.


%\bibliography{SPIN-BCS}

\begin{thebibliography}{39}%
\makeatletter
\providecommand \@ifxundefined [1]{%
 \@ifx{#1\undefined}
}%
\providecommand \@ifnum [1]{%
 \ifnum #1\expandafter \@firstoftwo
 \else \expandafter \@secondoftwo
 \fi
}%
\providecommand \@ifx [1]{%
 \ifx #1\expandafter \@firstoftwo
 \else \expandafter \@secondoftwo
 \fi
}%
\providecommand \natexlab [1]{#1}%
\providecommand \enquote  [1]{``#1''}%
\providecommand \bibnamefont  [1]{#1}%
\providecommand \bibfnamefont [1]{#1}%
\providecommand \citenamefont [1]{#1}%
\providecommand \href@noop [0]{\@secondoftwo}%
\providecommand \href [0]{\begingroup \@sanitize@url \@href}%
\providecommand \@href[1]{\@@startlink{#1}\@@href}%
\providecommand \@@href[1]{\endgroup#1\@@endlink}%
\providecommand \@sanitize@url [0]{\catcode `\\12\catcode `\$12\catcode
  `\&12\catcode `\#12\catcode `\^12\catcode `\_12\catcode `\%12\relax}%
\providecommand \@@startlink[1]{}%
\providecommand \@@endlink[0]{}%
\providecommand \url  [0]{\begingroup\@sanitize@url \@url }%
\providecommand \@url [1]{\endgroup\@href {#1}{\urlprefix }}%
\providecommand \urlprefix  [0]{URL }%
\providecommand \Eprint [0]{\href }%
\providecommand \doibase [0]{https://doi.org/}%
\providecommand \selectlanguage [0]{\@gobble}%
\providecommand \bibinfo  [0]{\@secondoftwo}%
\providecommand \bibfield  [0]{\@secondoftwo}%
\providecommand \translation [1]{[#1]}%
\providecommand \BibitemOpen [0]{}%
\providecommand \bibitemStop [0]{}%
\providecommand \bibitemNoStop [0]{.\EOS\space}%
\providecommand \EOS [0]{\spacefactor3000\relax}%
\providecommand \BibitemShut  [1]{\csname bibitem#1\endcsname}%
\let\auto@bib@innerbib\@empty
%</preamble>
\bibitem [{\citenamefont {Bardeen}\ \emph {et~al.}(1957)\citenamefont
  {Bardeen}, \citenamefont {Cooper},\ and\ \citenamefont
  {Schrieffer}}]{BCS1957}%
  \BibitemOpen
  \bibfield  {author} {\bibinfo {author} {\bibfnamefont {J.}~\bibnamefont
  {Bardeen}}, \bibinfo {author} {\bibfnamefont {L.~N.}\ \bibnamefont
  {Cooper}},\ and\ \bibinfo {author} {\bibfnamefont {J.~R.}\ \bibnamefont
  {Schrieffer}},\ }\bibfield  {title} {\bibinfo {title} {Theory of
  superconductivity},\ }\href@noop {} {\bibfield  {journal} {\bibinfo
  {journal} {Phys. Rev.}\ }\textbf {\bibinfo {volume} {108}},\ \bibinfo {pages}
  {1175} (\bibinfo {year} {1957})}\BibitemShut {NoStop}%
\bibitem [{\citenamefont {Bednorz}\ and\ \citenamefont
  {M\"{u}ller}(1986)}]{Muller1986}%
  \BibitemOpen
  \bibfield  {author} {\bibinfo {author} {\bibfnamefont {J.~G.}\ \bibnamefont
  {Bednorz}}\ and\ \bibinfo {author} {\bibfnamefont {K.~A.}\ \bibnamefont
  {M\"{u}ller}},\ }\bibfield  {title} {\bibinfo {title} {Possible high t$_c$
  superconductivity in the {Ba-La-Cu-O} system},\ }\href@noop {} {\bibfield
  {journal} {\bibinfo  {journal} {Z. Phys. B}\ }\textbf {\bibinfo {volume}
  {64}},\ \bibinfo {pages} {189} (\bibinfo {year} {1986})}\BibitemShut
  {NoStop}%
\bibitem [{\citenamefont {Emery}\ and\ \citenamefont
  {Kivelson}(1995)}]{Kivelson95}%
  \BibitemOpen
  \bibfield  {author} {\bibinfo {author} {\bibfnamefont {V.~J.}\ \bibnamefont
  {Emery}}\ and\ \bibinfo {author} {\bibfnamefont {S.~A.}\ \bibnamefont
  {Kivelson}},\ }\bibfield  {title} {\bibinfo {title} {Importance of phase
  fluctuation in superconductors with small superfluid density},\ }\href@noop
  {} {\bibfield  {journal} {\bibinfo  {journal} {Nature}\ }\textbf {\bibinfo
  {volume} {374}},\ \bibinfo {pages} {434} (\bibinfo {year}
  {1995})}\BibitemShut {NoStop}%
\bibitem [{\citenamefont {Hirsch}(2009)}]{Hirsch2009}%
  \BibitemOpen
  \bibfield  {author} {\bibinfo {author} {\bibfnamefont {J.}~\bibnamefont
  {Hirsch}},\ }\bibfield  {title} {\bibinfo {title} {{BCS} theory of
  superconductivity: it is time to question its validity},\ }\href@noop {}
  {\bibfield  {journal} {\bibinfo  {journal} {Physica Scripta}\ }\textbf
  {\bibinfo {volume} {80}},\ \bibinfo {pages} {035702} (\bibinfo {year}
  {2009})}\BibitemShut {NoStop}%
\bibitem [{\citenamefont {Koizumi}(2011)}]{Koizumi2011}%
  \BibitemOpen
  \bibfield  {author} {\bibinfo {author} {\bibfnamefont {H.}~\bibnamefont
  {Koizumi}},\ }\bibfield  {title} {\bibinfo {title} {Spin-vortex
  superconductivity},\ }\href@noop {} {\bibfield  {journal} {\bibinfo
  {journal} {J. Supercond. Nov. Magn.}\ }\textbf {\bibinfo {volume} {24}},\
  \bibinfo {pages} {1997} (\bibinfo {year} {2011})}\BibitemShut {NoStop}%
\bibitem [{\citenamefont {Hirsch}(2017)}]{Hirsch2017}%
  \BibitemOpen
  \bibfield  {author} {\bibinfo {author} {\bibfnamefont {J.~E.}\ \bibnamefont
  {Hirsch}},\ }\bibfield  {title} {\bibinfo {title} {Momentum of
  superconducting electrons and the explanation of the {Meissner} effect},\
  }\href@noop {} {\bibfield  {journal} {\bibinfo  {journal} {Phys Rev. B}\
  }\textbf {\bibinfo {volume} {95}},\ \bibinfo {pages} {014503} (\bibinfo
  {year} {2017})}\BibitemShut {NoStop}%
\bibitem [{\citenamefont {Hirsch}(2018)}]{Hirsch2018}%
  \BibitemOpen
  \bibfield  {author} {\bibinfo {author} {\bibfnamefont {J.~E.}\ \bibnamefont
  {Hirsch}},\ }\bibfield  {title} {\bibinfo {title} {Entropy generation and
  momentum transfer in the superconductor-normal and normal-superconductor
  phase transitions and the consistency of the conventional theory of
  superconductivity},\ }\href@noop {} {\bibfield  {journal} {\bibinfo
  {journal} {International Journal of Modern Physics B}\ }\textbf {\bibinfo
  {volume} {32}},\ \bibinfo {pages} {1850158} (\bibinfo {year}
  {2018})}\BibitemShut {NoStop}%
\bibitem [{\citenamefont {Hirsch}(2020)}]{Hirsch2020}%
  \BibitemOpen
  \bibfield  {author} {\bibinfo {author} {\bibfnamefont {J.~E.}\ \bibnamefont
  {Hirsch}},\ }\bibfield  {title} {\bibinfo {title} {Inconsistency of the
  conventional theory of superconductivity},\ }\href@noop {} {\bibfield
  {journal} {\bibinfo  {journal} {EPL}\ }\textbf {\bibinfo {volume} {130}},\
  \bibinfo {pages} {17006} (\bibinfo {year} {2020})}\BibitemShut {NoStop}%
\bibitem [{\citenamefont {Koizumi}(2020{\natexlab{a}})}]{koizumi2020b}%
  \BibitemOpen
  \bibfield  {author} {\bibinfo {author} {\bibfnamefont {H.}~\bibnamefont
  {Koizumi}},\ }\bibfield  {title} {\bibinfo {title} {Reversible
  superconducting-normal phase transition in a magnetic field and the existence
  of topologically-protected loop currents that appear and disappear without
  {Joule} heating},\ }\href@noop {} {\bibfield  {journal} {\bibinfo  {journal}
  {EPL}\ }\textbf {\bibinfo {volume} {131}},\ \bibinfo {pages} {37001}
  (\bibinfo {year} {2020}{\natexlab{a}})}\BibitemShut {NoStop}%
\bibitem [{\citenamefont {Keesom}\ and\ \citenamefont
  {Kok}(1934)}]{Keesom1934a}%
  \BibitemOpen
  \bibfield  {author} {\bibinfo {author} {\bibfnamefont {W.}~\bibnamefont
  {Keesom}}\ and\ \bibinfo {author} {\bibfnamefont {J.}~\bibnamefont {Kok}},\
  }\bibfield  {title} {\bibinfo {title} {Measurements of the latent heat of
  thallium connected with the transition, in a constant external magnetic
  field, from the supraconductive to the non-supraconductive state},\ }\href
  {https://doi.org/https://doi.org/10.1016/S0031-8914(34)90059-8} {\bibfield
  {journal} {\bibinfo  {journal} {Physica}\ }\textbf {\bibinfo {volume} {1}},\
  \bibinfo {pages} {503 } (\bibinfo {year} {1934})}\BibitemShut {NoStop}%
\bibitem [{\citenamefont {Keesom}\ and\ \citenamefont {{Van
  Laer}}(1936)}]{Keesom1934b}%
  \BibitemOpen
  \bibfield  {author} {\bibinfo {author} {\bibfnamefont {W.}~\bibnamefont
  {Keesom}}\ and\ \bibinfo {author} {\bibfnamefont {P.}~\bibnamefont {{Van
  Laer}}},\ }\bibfield  {title} {\bibinfo {title} {Measurements of the latent
  heat of tin in passing from the supraconductive to the non-supraconductive
  state},\ }\href
  {https://doi.org/https://doi.org/10.1016/S0031-8914(36)80002-0} {\bibfield
  {journal} {\bibinfo  {journal} {Physica}\ }\textbf {\bibinfo {volume} {3}},\
  \bibinfo {pages} {371 } (\bibinfo {year} {1936})}\BibitemShut {NoStop}%
\bibitem [{\citenamefont {Keesom}\ and\ \citenamefont {van
  Laer}(1937)}]{Keesom1937}%
  \BibitemOpen
  \bibfield  {author} {\bibinfo {author} {\bibfnamefont {W.}~\bibnamefont
  {Keesom}}\ and\ \bibinfo {author} {\bibfnamefont {P.}~\bibnamefont {van
  Laer}},\ }\bibfield  {title} {\bibinfo {title} {Measurements of the latent
  heat of tin while passing from the superconductive to the non-superconductive
  state at constant temperature},\ }\href
  {https://doi.org/https://doi.org/10.1016/S0031-8914(37)80081-6} {\bibfield
  {journal} {\bibinfo  {journal} {Physica}\ }\textbf {\bibinfo {volume} {4}},\
  \bibinfo {pages} {487 } (\bibinfo {year} {1937})}\BibitemShut {NoStop}%
\bibitem [{\citenamefont {van Laer}\ and\ \citenamefont
  {Keesom}(1938)}]{Keesom}%
  \BibitemOpen
  \bibfield  {author} {\bibinfo {author} {\bibfnamefont {P.~H.}\ \bibnamefont
  {van Laer}}\ and\ \bibinfo {author} {\bibfnamefont {W.~H.}\ \bibnamefont
  {Keesom}},\ }\bibfield  {title} {\bibinfo {title} {On the reversibility of
  the transition processs between the superconductive and the normal state},\
  }\href@noop {} {\bibfield  {journal} {\bibinfo  {journal} {Physica}\ }\textbf
  {\bibinfo {volume} {5}},\ \bibinfo {pages} {993} (\bibinfo {year}
  {1938})}\BibitemShut {NoStop}%
\bibitem [{\citenamefont {Hirsch}(2013)}]{Hirsch2013b}%
  \BibitemOpen
  \bibfield  {author} {\bibinfo {author} {\bibfnamefont {J.~E.}\ \bibnamefont
  {Hirsch}},\ }\bibfield  {title} {\bibinfo {title} {The {London} moment: what
  a rotating superconductor reveals about superconductivity},\ }\href
  {https://doi.org/10.1088/0031-8949/89/01/015806} {\bibfield  {journal}
  {\bibinfo  {journal} {Physica Scripta}\ }\textbf {\bibinfo {volume} {89}},\
  \bibinfo {pages} {015806} (\bibinfo {year} {2013})}\BibitemShut {NoStop}%
\bibitem [{\citenamefont {Koizumi}(2020{\natexlab{b}})}]{koizumi2021}%
  \BibitemOpen
  \bibfield  {author} {\bibinfo {author} {\bibfnamefont {H.}~\bibnamefont
  {Koizumi}},\ }\bibfield  {title} {\bibinfo {title} {London moment, {London's}
  superpotential, {Nambu-Goldstone} mode, and {Berry} connection from many-body
  wave functions},\ }\href@noop {} {\bibfield  {journal} {\bibinfo  {journal}
  {DOI: 10.1007/s10948-021-05827-9 J. Supercond. Nov. Magn.}\ } (\bibinfo
  {year} {2020}{\natexlab{b}})},\ \Eprint {https://arxiv.org/abs/2011.10701}
  {arXiv:2011.10701 [cond-mat.supr-con]} \BibitemShut {NoStop}%
\bibitem [{\citenamefont {Hildebrandt}(1964)}]{Hildebrandt1964}%
  \BibitemOpen
  \bibfield  {author} {\bibinfo {author} {\bibfnamefont {A.~F.}\ \bibnamefont
  {Hildebrandt}},\ }\bibfield  {title} {\bibinfo {title} {Magnetic field of a
  rotating superconductor},\ }\href
  {https://doi.org/10.1103/PhysRevLett.12.190} {\bibfield  {journal} {\bibinfo
  {journal} {Phys. Rev. Lett.}\ }\textbf {\bibinfo {volume} {12}},\ \bibinfo
  {pages} {190} (\bibinfo {year} {1964})}\BibitemShut {NoStop}%
\bibitem [{\citenamefont {Zimmerman}\ and\ \citenamefont
  {Mercereau}(1965)}]{Zimmerman1965}%
  \BibitemOpen
  \bibfield  {author} {\bibinfo {author} {\bibfnamefont {J.~E.}\ \bibnamefont
  {Zimmerman}}\ and\ \bibinfo {author} {\bibfnamefont {J.~E.}\ \bibnamefont
  {Mercereau}},\ }\bibfield  {title} {\bibinfo {title} {Compton wavelength of
  superconducting electrons},\ }\href
  {https://doi.org/10.1103/PhysRevLett.14.887} {\bibfield  {journal} {\bibinfo
  {journal} {Phys. Rev. Lett.}\ }\textbf {\bibinfo {volume} {14}},\ \bibinfo
  {pages} {887} (\bibinfo {year} {1965})}\BibitemShut {NoStop}%
\bibitem [{\citenamefont {Brickman}(1969)}]{Brickman1969}%
  \BibitemOpen
  \bibfield  {author} {\bibinfo {author} {\bibfnamefont {N.~F.}\ \bibnamefont
  {Brickman}},\ }\bibfield  {title} {\bibinfo {title} {Rotating
  superconductors},\ }\href {https://doi.org/10.1103/PhysRev.184.460}
  {\bibfield  {journal} {\bibinfo  {journal} {Phys. Rev.}\ }\textbf {\bibinfo
  {volume} {184}},\ \bibinfo {pages} {460} (\bibinfo {year}
  {1969})}\BibitemShut {NoStop}%
\bibitem [{\citenamefont {Tate}\ \emph {et~al.}(1989)\citenamefont {Tate},
  \citenamefont {Cabrera}, \citenamefont {Felch},\ and\ \citenamefont
  {Anderson}}]{Tate1989}%
  \BibitemOpen
  \bibfield  {author} {\bibinfo {author} {\bibfnamefont {J.}~\bibnamefont
  {Tate}}, \bibinfo {author} {\bibfnamefont {B.}~\bibnamefont {Cabrera}},
  \bibinfo {author} {\bibfnamefont {S.~B.}\ \bibnamefont {Felch}},\ and\
  \bibinfo {author} {\bibfnamefont {J.~T.}\ \bibnamefont {Anderson}},\
  }\bibfield  {title} {\bibinfo {title} {Precise determination of the
  {Cooper}-pair mass},\ }\href {https://doi.org/10.1103/PhysRevLett.62.845}
  {\bibfield  {journal} {\bibinfo  {journal} {Phys. Rev. Lett.}\ }\textbf
  {\bibinfo {volume} {62}},\ \bibinfo {pages} {845} (\bibinfo {year}
  {1989})}\BibitemShut {NoStop}%
\bibitem [{\citenamefont {Tate}\ \emph {et~al.}(1990)\citenamefont {Tate},
  \citenamefont {Felch},\ and\ \citenamefont {Cabrera}}]{Tate1990}%
  \BibitemOpen
  \bibfield  {author} {\bibinfo {author} {\bibfnamefont {J.}~\bibnamefont
  {Tate}}, \bibinfo {author} {\bibfnamefont {S.~B.}\ \bibnamefont {Felch}},\
  and\ \bibinfo {author} {\bibfnamefont {B.}~\bibnamefont {Cabrera}},\
  }\bibfield  {title} {\bibinfo {title} {Determination of the {Cooper}-pair
  mass in niobium},\ }\href {https://doi.org/10.1103/PhysRevB.42.7885}
  {\bibfield  {journal} {\bibinfo  {journal} {Phys. Rev. B}\ }\textbf {\bibinfo
  {volume} {42}},\ \bibinfo {pages} {7885} (\bibinfo {year}
  {1990})}\BibitemShut {NoStop}%
\bibitem [{\citenamefont {Verheijen}\ \emph
  {et~al.}(1990{\natexlab{a}})\citenamefont {Verheijen}, \citenamefont {{van
  Ruitenbeek}}, \citenamefont {{de Bruyn Ouboter}},\ and\ \citenamefont {{de
  Jongh}}}]{VERHEIJEN1990a}%
  \BibitemOpen
  \bibfield  {author} {\bibinfo {author} {\bibfnamefont {A.}~\bibnamefont
  {Verheijen}}, \bibinfo {author} {\bibfnamefont {J.}~\bibnamefont {{van
  Ruitenbeek}}}, \bibinfo {author} {\bibfnamefont {R.}~\bibnamefont {{de Bruyn
  Ouboter}}},\ and\ \bibinfo {author} {\bibfnamefont {L.}~\bibnamefont {{de
  Jongh}}},\ }\bibfield  {title} {\bibinfo {title} {The {London} moment for
  high temperature superconductors},\ }\href
  {https://doi.org/https://doi.org/10.1016/S0921-4526(09)80176-2} {\bibfield
  {journal} {\bibinfo  {journal} {Physica B: Condensed Matter}\ }\textbf
  {\bibinfo {volume} {165-166}},\ \bibinfo {pages} {1181 } (\bibinfo {year}
  {1990}{\natexlab{a}})},\ \bibinfo {note} {lT-19}\BibitemShut {NoStop}%
\bibitem [{\citenamefont {Verheijen}\ \emph
  {et~al.}(1990{\natexlab{b}})\citenamefont {Verheijen}, \citenamefont {van
  Ruitenbeek}, \citenamefont {de~Bruyn~Ouboter},\ and\ \citenamefont
  {de~Jongh}}]{Verheijen1990}%
  \BibitemOpen
  \bibfield  {author} {\bibinfo {author} {\bibfnamefont {A.~A.}\ \bibnamefont
  {Verheijen}}, \bibinfo {author} {\bibfnamefont {J.~M.}\ \bibnamefont {van
  Ruitenbeek}}, \bibinfo {author} {\bibfnamefont {R.}~\bibnamefont
  {de~Bruyn~Ouboter}},\ and\ \bibinfo {author} {\bibfnamefont {L.~J.}\
  \bibnamefont {de~Jongh}},\ }\bibfield  {title} {\bibinfo {title} {Measurement
  of the {London} moment in two high-temperature superconductors},\ }\href
  {https://doi.org/10.1038/345418a0} {\bibfield  {journal} {\bibinfo  {journal}
  {Nature}\ }\textbf {\bibinfo {volume} {345}},\ \bibinfo {pages} {418}
  (\bibinfo {year} {1990}{\natexlab{b}})}\BibitemShut {NoStop}%
\bibitem [{\citenamefont {Sanzari}\ \emph {et~al.}(1996)\citenamefont
  {Sanzari}, \citenamefont {Cui},\ and\ \citenamefont
  {Karwacki}}]{Sanzari1996}%
  \BibitemOpen
  \bibfield  {author} {\bibinfo {author} {\bibfnamefont {M.~A.}\ \bibnamefont
  {Sanzari}}, \bibinfo {author} {\bibfnamefont {H.~L.}\ \bibnamefont {Cui}},\
  and\ \bibinfo {author} {\bibfnamefont {F.}~\bibnamefont {Karwacki}},\
  }\bibfield  {title} {\bibinfo {title} {London moment for heavy-fermion
  superconductors},\ }\href {https://doi.org/10.1063/1.116622} {\bibfield
  {journal} {\bibinfo  {journal} {Applied Physics Letters}\ }\textbf {\bibinfo
  {volume} {68}},\ \bibinfo {pages} {3802} (\bibinfo {year}
  {1996})}\BibitemShut {NoStop}%
\bibitem [{\citenamefont {Gor'kov}(1959)}]{Gorkov1959}%
  \BibitemOpen
  \bibfield  {author} {\bibinfo {author} {\bibfnamefont {L.~P.}\ \bibnamefont
  {Gor'kov}},\ }\bibfield  {title} {\bibinfo {title} {Microscopic derivation of
  the {Ginzburg-Landau} equations in the theory of superconductivity},\
  }\href@noop {} {\bibfield  {journal} {\bibinfo  {journal} {Sov. Phys. JETP}\
  }\textbf {\bibinfo {volume} {36}},\ \bibinfo {pages} {1364} (\bibinfo {year}
  {1959})}\BibitemShut {NoStop}%
\bibitem [{\citenamefont {Koizumi}(2020{\natexlab{c}})}]{koizumi2019}%
  \BibitemOpen
  \bibfield  {author} {\bibinfo {author} {\bibfnamefont {H.}~\bibnamefont
  {Koizumi}},\ }\bibfield  {title} {\bibinfo {title} {Explanation of
  superfluidity using the {Berry} connection for many-body wave functions},\
  }\href@noop {} {\bibfield  {journal} {\bibinfo  {journal}
  {J.~Supercond.~Nov.~Magn.}\ }\textbf {\bibinfo {volume} {33}},\ \bibinfo
  {pages} {1697} (\bibinfo {year} {2020}{\natexlab{c}})}\BibitemShut {NoStop}%
\bibitem [{\citenamefont {Schmid}(1983)}]{Schmid1983}%
  \BibitemOpen
  \bibfield  {author} {\bibinfo {author} {\bibfnamefont {A.}~\bibnamefont
  {Schmid}},\ }\bibfield  {title} {\bibinfo {title} {Diffusion and localization
  in a dissipative quantum system},\ }\href
  {https://doi.org/10.1103/PhysRevLett.51.1506} {\bibfield  {journal} {\bibinfo
   {journal} {Phys. Rev. Lett.}\ }\textbf {\bibinfo {volume} {51}},\ \bibinfo
  {pages} {1506} (\bibinfo {year} {1983})}\BibitemShut {NoStop}%
\bibitem [{\citenamefont {Hakonen}\ and\ \citenamefont
  {Sonin}(2021)}]{PhysRevX2021b}%
  \BibitemOpen
  \bibfield  {author} {\bibinfo {author} {\bibfnamefont {P.~J.}\ \bibnamefont
  {Hakonen}}\ and\ \bibinfo {author} {\bibfnamefont {E.~B.}\ \bibnamefont
  {Sonin}},\ }\bibfield  {title} {\bibinfo {title} {Comment on ``absence of a
  dissipative quantum phase transition in {Josephson} junctions''},\ }\href
  {https://doi.org/10.1103/PhysRevX.11.018001} {\bibfield  {journal} {\bibinfo
  {journal} {Phys. Rev. X}\ }\textbf {\bibinfo {volume} {11}},\ \bibinfo
  {pages} {018001} (\bibinfo {year} {2021})}\BibitemShut {NoStop}%
\bibitem [{\citenamefont {Murani}\ \emph {et~al.}(2020)\citenamefont {Murani},
  \citenamefont {Bourlet}, \citenamefont {le~Sueur}, \citenamefont {Portier},
  \citenamefont {Altimiras}, \citenamefont {Esteve}, \citenamefont {Grabert},
  \citenamefont {Stockburger}, \citenamefont {Ankerhold},\ and\ \citenamefont
  {Joyez}}]{PhysRevX2021a}%
  \BibitemOpen
  \bibfield  {author} {\bibinfo {author} {\bibfnamefont {A.}~\bibnamefont
  {Murani}}, \bibinfo {author} {\bibfnamefont {N.}~\bibnamefont {Bourlet}},
  \bibinfo {author} {\bibfnamefont {H.}~\bibnamefont {le~Sueur}}, \bibinfo
  {author} {\bibfnamefont {F.}~\bibnamefont {Portier}}, \bibinfo {author}
  {\bibfnamefont {C.}~\bibnamefont {Altimiras}}, \bibinfo {author}
  {\bibfnamefont {D.}~\bibnamefont {Esteve}}, \bibinfo {author} {\bibfnamefont
  {H.}~\bibnamefont {Grabert}}, \bibinfo {author} {\bibfnamefont
  {J.}~\bibnamefont {Stockburger}}, \bibinfo {author} {\bibfnamefont
  {J.}~\bibnamefont {Ankerhold}},\ and\ \bibinfo {author} {\bibfnamefont
  {P.}~\bibnamefont {Joyez}},\ }\bibfield  {title} {\bibinfo {title} {Absence
  of a dissipative quantum phase transition in josephson junctions},\ }\href
  {https://doi.org/10.1103/PhysRevX.10.021003} {\bibfield  {journal} {\bibinfo
  {journal} {Phys. Rev. X}\ }\textbf {\bibinfo {volume} {10}},\ \bibinfo
  {pages} {021003} (\bibinfo {year} {2020})}\BibitemShut {NoStop}%
\bibitem [{\citenamefont {Murani}\ \emph {et~al.}(2021)\citenamefont {Murani},
  \citenamefont {Bourlet}, \citenamefont {le~Sueur}, \citenamefont {Portier},
  \citenamefont {Altimiras}, \citenamefont {Esteve}, \citenamefont {Grabert},
  \citenamefont {Stockburger}, \citenamefont {Ankerhold},\ and\ \citenamefont
  {Joyez}}]{PhysRevX2021c}%
  \BibitemOpen
  \bibfield  {author} {\bibinfo {author} {\bibfnamefont {A.}~\bibnamefont
  {Murani}}, \bibinfo {author} {\bibfnamefont {N.}~\bibnamefont {Bourlet}},
  \bibinfo {author} {\bibfnamefont {H.}~\bibnamefont {le~Sueur}}, \bibinfo
  {author} {\bibfnamefont {F.}~\bibnamefont {Portier}}, \bibinfo {author}
  {\bibfnamefont {C.}~\bibnamefont {Altimiras}}, \bibinfo {author}
  {\bibfnamefont {D.}~\bibnamefont {Esteve}}, \bibinfo {author} {\bibfnamefont
  {H.}~\bibnamefont {Grabert}}, \bibinfo {author} {\bibfnamefont
  {J.}~\bibnamefont {Stockburger}}, \bibinfo {author} {\bibfnamefont
  {J.}~\bibnamefont {Ankerhold}},\ and\ \bibinfo {author} {\bibfnamefont
  {P.}~\bibnamefont {Joyez}},\ }\bibfield  {title} {\bibinfo {title} {Reply to
  ``comment on `absence of a dissipative quantum phase transition in
  {Josephson} junctions'''},\ }\href
  {https://doi.org/10.1103/PhysRevX.11.018002} {\bibfield  {journal} {\bibinfo
  {journal} {Phys. Rev. X}\ }\textbf {\bibinfo {volume} {11}},\ \bibinfo
  {pages} {018002} (\bibinfo {year} {2021})}\BibitemShut {NoStop}%
\bibitem [{\citenamefont {Devoret}(1997)}]{Devoret1997}%
  \BibitemOpen
  \bibfield  {author} {\bibinfo {author} {\bibfnamefont {M.}~\bibnamefont
  {Devoret}},\ }\bibfield  {title} {\bibinfo {title} {Quantum fluctuations in
  electrical circuits},\ }in\ \href@noop {} {\emph {\bibinfo {booktitle}
  {Quantum Fluctuations: Les Houches Session LXIII}}},\ \bibinfo {editor}
  {edited by\ \bibinfo {editor} {\bibfnamefont {S.~Z.-J.~J.}\ \bibnamefont
  {Reynaud}, \bibfnamefont {S.;~Giacobino}}}\ (\bibinfo  {publisher}
  {Elsevier},\ \bibinfo {year} {1997})\BibitemShut {NoStop}%
\bibitem [{\citenamefont {Gu}\ \emph {et~al.}(2017)\citenamefont {Gu},
  \citenamefont {Kochum}, \citenamefont {Miranowicz}, \citenamefont {Liu},\
  and\ \citenamefont {Nori}}]{Nori2017}%
  \BibitemOpen
  \bibfield  {author} {\bibinfo {author} {\bibfnamefont {X.}~\bibnamefont
  {Gu}}, \bibinfo {author} {\bibfnamefont {A.~F.}\ \bibnamefont {Kochum}},
  \bibinfo {author} {\bibfnamefont {A.}~\bibnamefont {Miranowicz}}, \bibinfo
  {author} {\bibfnamefont {Y.}~\bibnamefont {Liu}},\ and\ \bibinfo {author}
  {\bibfnamefont {F.}~\bibnamefont {Nori}},\ }\href@noop {} {\bibfield
  {journal} {\bibinfo  {journal} {Phys. Report}\ }\textbf {\bibinfo {volume}
  {718-719}},\ \bibinfo {pages} {1} (\bibinfo {year} {2017})}\BibitemShut
  {NoStop}%
\bibitem [{\citenamefont {Josephson}(1962)}]{Josephson62}%
  \BibitemOpen
  \bibfield  {author} {\bibinfo {author} {\bibfnamefont {B.~D.}\ \bibnamefont
  {Josephson}},\ }\bibfield  {title} {\bibinfo {title} {Possible new effects in
  superconductive tunneling},\ }\href@noop {} {\bibfield  {journal} {\bibinfo
  {journal} {Phys. Lett.}\ }\textbf {\bibinfo {volume} {1}},\ \bibinfo {pages}
  {251} (\bibinfo {year} {1962})}\BibitemShut {NoStop}%
\bibitem [{\citenamefont {Cohen}\ \emph {et~al.}(1962)\citenamefont {Cohen},
  \citenamefont {Falicov},\ and\ \citenamefont {Phillips}}]{Cohen1962}%
  \BibitemOpen
  \bibfield  {author} {\bibinfo {author} {\bibfnamefont {M.~H.}\ \bibnamefont
  {Cohen}}, \bibinfo {author} {\bibfnamefont {L.~M.}\ \bibnamefont {Falicov}},\
  and\ \bibinfo {author} {\bibfnamefont {J.~C.}\ \bibnamefont {Phillips}},\
  }\bibfield  {title} {\bibinfo {title} {Superconductive tunneling},\ }\href
  {https://doi.org/10.1103/PhysRevLett.8.316} {\bibfield  {journal} {\bibinfo
  {journal} {Phys. Rev. Lett.}\ }\textbf {\bibinfo {volume} {8}},\ \bibinfo
  {pages} {316} (\bibinfo {year} {1962})}\BibitemShut {NoStop}%
\bibitem [{\citenamefont {Ferrell}\ and\ \citenamefont
  {Prange}(1963)}]{Ferrell1963}%
  \BibitemOpen
  \bibfield  {author} {\bibinfo {author} {\bibfnamefont {R.~A.}\ \bibnamefont
  {Ferrell}}\ and\ \bibinfo {author} {\bibfnamefont {R.~E.}\ \bibnamefont
  {Prange}},\ }\bibfield  {title} {\bibinfo {title} {Self-field limiting of
  josephson tunneling of superconducting electron pairs},\ }\href
  {https://doi.org/10.1103/PhysRevLett.10.479} {\bibfield  {journal} {\bibinfo
  {journal} {Phys. Rev. Lett.}\ }\textbf {\bibinfo {volume} {10}},\ \bibinfo
  {pages} {479} (\bibinfo {year} {1963})}\BibitemShut {NoStop}%
\bibitem [{\citenamefont {Josephson}(1965)}]{Josephson1965}%
  \BibitemOpen
  \bibfield  {author} {\bibinfo {author} {\bibfnamefont {B.~D.}\ \bibnamefont
  {Josephson}},\ }\bibfield  {title} {\bibinfo {title} {Supercurrents through
  barriers},\ }\href@noop {} {\bibfield  {journal} {\bibinfo  {journal} {Adv.
  Phys.}\ }\textbf {\bibinfo {volume} {14}},\ \bibinfo {pages} {419} (\bibinfo
  {year} {1965})}\BibitemShut {NoStop}%
\bibitem [{\citenamefont {Josephson}(1969)}]{Josephson1969}%
  \BibitemOpen
  \bibfield  {author} {\bibinfo {author} {\bibfnamefont {B.~D.}\ \bibnamefont
  {Josephson}},\ }\bibfield  {title} {\bibinfo {title} {Weakly coupled
  superconductos},\ }in\ \href@noop {} {\emph {\bibinfo {booktitle}
  {Superconductivity}}},\ Vol.~\bibinfo {volume} {1},\ \bibinfo {editor}
  {edited by\ \bibinfo {editor} {\bibfnamefont {R.~D.}\ \bibnamefont {Parks}}}\
  (\bibinfo  {publisher} {Marcel Dekker, inc.},\ \bibinfo {address} {New
  York},\ \bibinfo {year} {1969})\ Chap.~\bibinfo {chapter} {9}, p.\ \bibinfo
  {pages} {423}\BibitemShut {NoStop}%
\bibitem [{\citenamefont {Anderson}(1964)}]{Anderson64}%
  \BibitemOpen
  \bibfield  {author} {\bibinfo {author} {\bibfnamefont {P.~W.}\ \bibnamefont
  {Anderson}},\ }\bibfield  {title} {\bibinfo {title} {Special effects in
  superconductivity},\ }in\ \href@noop {} {\emph {\bibinfo {booktitle}
  {Lectures on the many-body problems}}},\ Vol.~\bibinfo {volume} {2},\
  \bibinfo {editor} {edited by\ \bibinfo {editor} {\bibfnamefont {E.~R.}\
  \bibnamefont {Caianiello}}}\ (\bibinfo  {publisher} {Academic Press},\
  \bibinfo {year} {1964})\ pp.\ \bibinfo {pages} {113--135}\BibitemShut
  {NoStop}%
\bibitem [{\citenamefont {Shapiro}(1963)}]{Shapiro63}%
  \BibitemOpen
  \bibfield  {author} {\bibinfo {author} {\bibfnamefont {S.}~\bibnamefont
  {Shapiro}},\ }\bibfield  {title} {\bibinfo {title} {Josephson currents in
  superconducting tunneling: the effect of microwaves and other observations},\
  }\href@noop {} {\bibfield  {journal} {\bibinfo  {journal} {Phys. Rev. Lett.}\
  }\textbf {\bibinfo {volume} {11}},\ \bibinfo {pages} {80} (\bibinfo {year}
  {1963})}\BibitemShut {NoStop}%
\bibitem [{\citenamefont {Tinkham}(1996)}]{TinkhamText}%
  \BibitemOpen
  \bibfield  {author} {\bibinfo {author} {\bibfnamefont {M.}~\bibnamefont
  {Tinkham}},\ }\href@noop {} {\emph {\bibinfo {title} {Introduction to
  superconductivity}}},\ \bibinfo {edition} {2nd}\ ed.\ (\bibinfo  {publisher}
  {MacGraw-Hill},\ \bibinfo {address} {USA},\ \bibinfo {year}
  {1996})\BibitemShut {NoStop}%
\end{thebibliography}

\end{document}